\documentclass[aps,prl,twocolumn,tightenlines,showpacs,aps,amsmath,amssymb,nofootinbib]{revtex4-1}

\usepackage{latexsym}
\usepackage{graphicx}
\usepackage{subfigure}
\usepackage{cases}
\usepackage{hyperref}
\usepackage{amssymb}
\usepackage{bm}
\usepackage{natbib}
\usepackage{amsmath}
\usepackage{slashed}
\usepackage{braket}
\usepackage{color}

\newcommand{\be}{\begin{equation}}
\newcommand{\ee}{\end{equation}}
\newcommand{\ba}{\begin{eqnarray}}
\newcommand{\ea}{\end{eqnarray}}
\def\bal{\begin{align}}
\def\eal{\end{align}}
\def\bald{\begin{aligned}}
\def\eald{\end{aligned}}

\begin{document}

\title{Two-flavor adjoint QCD}

\date{\today}

\author{Mohamed M. Anber}
\email[]{manber@lclark.edu}
\affiliation{Department of Physics, Lewis $\&$ Clark College, Portland, OR 97219, USA}
\author{Erich Poppitz}
\email[]{poppitz@physics.utoronto.ca}
\affiliation{Department of Physics, University of Toronto, Toronto, ON M5S 1A7, Canada}
%=========================================================================================

\begin{abstract}
We study four dimensional $SU(2)$ Yang-Mills theory with two  massless adjoint Weyl fermions.  When compactified on a spatial circle  of size $L$ much smaller than the strong-coupling scale, this theory can be solved by weak-coupling nonperturbative  semiclassical methods. We study the possible realizations of symmetries in the $\mathbb R^4$ limit  and find that all continuous and discrete   $0$-form and $1$-form 't Hooft anomaly matching conditions are saturated by a symmetry realization and massless spectrum identical to that found in the small-$L$ limit, with only a single massless flavor-doublet fermion in the infrared. This observation raises the possibility that the class of theories which undergo no phase transition between the analytically-solvable small-size circle and strongly-coupled infinite-size circle is  larger than previously thought, and offers new challenges for lattice studies.
\end{abstract}
%\pacs{
   %   }

%=========================================================================================

\maketitle

{\flushleft\textbf{Introduction:}} The solution of general four-dimensional strongly coupled gauge theories remains an elusive goal. Assuming large amounts of symmetry, e.g. extended supersymmetry,  helps make progress, but the resulting theories are often quite distinct from the ones describing the known interactions of elementary particles, or from  various conjectured extensions of the Standard Model. Putting gauge theories on a space-time lattice has proven  tremendously useful, but is a subject of severe technical (or conceptual) difficulties, especially for  theories with global (or gauged) chiral symmetries. 

Thus, any new handle to study the nonperturbative behavior of four-dimensional gauge theories should be met with excitement and thoroughly examined. 
In this paper we do this by combining two relatively recent interesting developments. 

The first is the now ten-year old observation \cite{Unsal:2007jx} (reviewed in \cite{Dunne:2016nmc})   that compactification of large classes of four-dimensional gauge theories on a circle allows for controlled nonperturbative studies of their dynamics. The control parameter is the circle size $L$, which, when taken small compared to the strong coupling scale of the theory, allows for controlled studies of the ground state symmetries and spectrum. Whether the  symmetry realization and spectrum change continuously in the large-$L$ limit of physical interest, however, remains a difficult question, not yet answered for general classes of theories.

The second development  is the more recent discovery of novel  anomaly matching conditions, in the spirit of 't Hooft, \cite{Frishman:1980dq,Coleman:1982yg} and references therein, but involving higher-form (here: discrete) symmetries \cite{Gaiotto:2014kfa,Gaiotto:2017yup}. 

In this paper, we find a novel solution to the  't Hooft anomaly matching conditions for the nonsupersymmetric theory outlined in the abstract; more detail follows below. Remarkably, we find that the symmetry realization and massless spectrum in the calculable small-$L$ regime is identical to that of the solution of anomaly matching on $\mathbb R^4$. 

While the theory we study is just one example,  we hope to convince the reader that 
 the matching is not completely trivial and that it points to the possible existence of larger classes of theories where a calculable regime---achieved by introducing a control parameter (here: $L$)---is continuously connected to the regime of physical interest. A deeper understanding of the correspondence between the two regimes is, clearly, very desirable.

%%%%%%%%%%%%%%%%
{\flushleft\textbf{Theory and symmetries:}} We consider $SU(N_c=2)$ Yang-Mills theory on $\mathbb R^4$   with $n_f=2$ massless adjoint Weyl fermions $\lambda$. The classical global symmetries of the theory are the continuous   chiral (flavor) $SU_f(2)$ and axial $U(1)$. The continuous $U(1)$ axial group  is anomalous and is broken to a $\mathbb Z^{d\chi}_{4nf}$ discrete chiral symmetry, as can be  seen from the action of a $U(1)$ transformation on the 't Hooft vertex due to  a  Belavin-Polyakov-Schwarz-Tyupkin (BPST) instanton. The chiral $\mathbb Z^{d\chi}_{4nf}$ is a $0$-form symmetry since it acts on  local field operators. In addition, the theory can be probed by fundamental representation Wilson line operators, $W=\mbox{tr}\exp\left[\oint_C a\right]$, where  $a$ is the gauge field and $C$ is a noncontractible closed path (it helps to think of the theory defined on a  large $\mathbb T^4$). Such operators transform under the $\mathbb Z_2^C$ $1$-form center symmetry.  Thus,  the full global  (non-spacetime) symmetry of the theory   is $G=  {SU_f(2)\times \mathbb Z^{d\chi}_{4nf}} \times \mathbb Z_2^C$ (note that  $\mathbb Z_2$ fermion number       and   the center of $SU_f(2)$ also act as elements of $Z^{d\chi}_{4nf}$).

{\flushleft\textbf{Anomaly matching:}} Given a theory with a global symmetry $G$, it may be interesting to turn on  background gauge fields of $G$. If $G$ has a 't Hooft anomaly, $G$ gauge invariance cannot be maintained. The obstruction to gauging the global symmetry, which is usually easily seen in the free ultraviolet (UV) theory, is  renormalization group invariant and must be matched by the infrared (IR) dynamics of the theory. This ``anomaly matching" can be a powerful tool to put constraints on the theory in its strongly coupled regime. 

In the following, we study the 't Hooft anomalies of the two-flavor QCD(adj). We examine the matching in the zero-temperature phase of the theory  on $\mathbb R^4$ and find a novel solution to the anomaly matching conditions. This solution realizes the symmetries on $\mathbb R^4$ in the same way that they are known to be realized upon compactification on  $\mathbb R^3\times \mathbb S^1_L$  at  $L \Lambda_{QCD}\ll 1$  ($\Lambda_{QCD}$ is the strong-coupling scale of the theory), where the theory was  solved using semiclassical methods  \cite{Unsal:2007jx}. 

It has been usually thought that in the non-supersymmetric case of $n_f >1$ massless adjoint fermions there is a phase transition, upon increasing $L$ past $1/\Lambda_{QCD}$, associated either with the breaking of $SU(n_f)$ flavor symmetry, for small $n_f$, or with the restoration of the discrete chiral symmetry, for values of $n_f$ such that the theory becomes conformal\footnote{A  conformal field theory has no scale, and hence, the discrete chiral symmetry cannot be broken.} on $\mathbb R^4$. However, as our anomaly matching example shows,  continuity between the small $\mathbb S^1_L$ and the $\mathbb R^4$ limits may be a feature more general than the known cases of supersymmetric Yang-Mills theory or deformed Yang-Mills theory---where either formal arguments or a large body of evidence in favor of continuity exist.

%%%%%
{\flushleft\textbf{The zero temperature phase of the theory on ${\mathbb R}^4$:}} We propose that the theory is in a confined phase with unbroken $SU_f(2)$  and broken discrete chiral symmetry $\mathbb Z^{d\chi}_8 \rightarrow \mathbb Z^{d\chi}_4$. Therefore, the theory admits two vacua which transform into each other under $\mathbb Z^{d\chi}_8$. In the following we support our claim by showing that the proposed IR spectrum of the theory saturates the UV 't Hooft anomalies. 

In order to examine the breaking of the discrete chiral symmetry,  consider the  four-fermion operator, which is a singlet under the flavor $SU_f(2)$ and the gauge symmetry, and transforms non-trivially under $\mathbb Z^{d\chi}_8$:
\begin{eqnarray}\label{fourfermi}{\cal O}^{(1)}\equiv \left(\epsilon_{\alpha\beta}\lambda^{\alpha a}_{i}\lambda_{j}^{\beta a}\right)\left(\epsilon_{\alpha'\beta'}\lambda^{\alpha' a'}_{i'}\lambda^{\beta' a'}_{j'}\right)\epsilon^{i i'}\epsilon^{j j'}\,,
\end{eqnarray}
where repeated indices are summed over ($\alpha,\beta=1,2$ denote $SL(2,C)$ Lorentz indices, $i,j=1,2$ are flavor indices, while   $a,b=1,2,3$ are reserved for  color). It is trivial to see that ${\cal O}^{(1)}$ acquires a phase $e^{i\pi}$ under $\mathbb Z_8^{d\chi}$, and hence, this operator can be used to probe the breaking of this symmetry, e.g., in a lattice setup. 

Our proposal for an IR behavior of the $\mathbb R^4$ theory  is that at a scale of order $\Lambda_{QCD}$ the four-fermion operator (\ref{fourfermi}) acquires an expectation value breaking the discrete chiral symmetry, $\mathbb Z_8^{d\chi}\rightarrow \mathbb Z_4^{d\chi}$, but preserving $SU_f(2)$. The fermion bilinear, $\epsilon_{\alpha\beta}\lambda^{\alpha a}_{i}\lambda_{j}^{\beta a}$,  usually thought responsible for the breaking of $SU_f(2) \rightarrow SO_f(2)$, is assumed to vanish---as  it does in the semiclassical small-circle limit.

The second part of our proposal concerns the massless spectrum of the theory. It consists of a single massless hadron, composed of three adjoint fields, with an interpolating gauge invariant local operator 
 \begin{eqnarray}
\label{hadron}
{\cal O}_{[ij]k}^{(2)\,\gamma}&\equiv&\epsilon_{\alpha\beta}\lambda_{[i}^{\alpha\,a}\lambda_{j]}^{\beta\,b}\lambda^{\gamma c}_{k}\epsilon^{abc} . \end{eqnarray}
As indicated above,  the operator is antisymmetric in the indices $i,j$, hence the massless hadron (\ref{hadron}) transforms as a fundamental under $SU_f(2)$ and  a Weyl spinor under the Lorentz group. It also carries charge $3$ under $\mathbb Z_{8}^{d\chi}$. 
The charges of our order parameter ${\cal O}^{(1)}$  and massless hadron ${\cal O}^{(2)}$  are summarized in  Table \ref{charge table}, where we also list the charges of the UV adjoint fields $\lambda_i$. We next argue that the massless hadron (\ref{hadron}) saturates the 't Hooft anomalies for all global symmetries.

\begin{table}
\begin{center}
\begin{tabular}{|c|c|c|c|c|}
\hline
& $SU_{c}(2)$ & $SU_{f}(2)$ & $\mathbb Z_8^{d\chi}$ &  $\mathbb Z_4^{d\chi}$\\
\hline
$\lambda_i^\alpha$ & adj & $\Box$ &$1$ & $1$\\
\hline
${\cal O}^{(1)}$& $1$ & $1$ & $4$ & $4\equiv 0$\\
\hline
${\cal O}_{[ij]k}^{(2)\,\gamma}$& $1$ & $\Box$ & $3$ & $3$\\
\hline
\end{tabular}
\caption{ \label{charge table} The charges of the elementary and composite fields under the symmetries of the theory. }
\end{center}
\end{table}

  We start with the only continuous 't Hooft anomaly, $\left[SU_f(2)\right]^3$. There is  an odd number of  $SU_f(2)$ fundamentals ($N_c^2-1$$=$$3$) in the UV, and thus,  $SU_f(2)$ has a Witten anomaly. To saturate the anomaly in the IR  the theory should have an odd number of massless fermions charged under $SU_f(2)$. Clearly the assumed  single massless color-singlet fermion (\ref{hadron}) will do the trick. 

The remaining non vanishing 't Hooft anomalies all involve discrete $0$-form and $1$-form symmetries.
We begin with the anomalies involving $0$-form discrete symmetries discussed some time ago in \cite{Csaki:1997aw}. Following  their classification, we consider first the more constraining ``type-I" discrete anomalies  $\mathbb Z_{4}^{d\chi} \left[SU_f(2)\right]^2 $ and  $\mathbb Z_{4}^{d\chi}\left[{\cal G}\right]^2 $, where ${\cal G}$ denotes a background gravitational field. Notice that it suffices to consider the unbroken part of the discrete chiral symmetry (although, from Table \ref{charge table}, this makes no difference as the charge assignments are identical, see also footnote \ref{six}).

To compute the $\mathbb Z^{d\chi}_{4}\left[SU_f(2)\right]^2$ anomaly,  consider  an $SU_f(2)$ BPST instanton background and note that the number of fermionic zero modes   is $N_c^2-1=3$, the number of $SU_f(2)$ fundamentals in the UV (we remind the reader that we are counting the zero modes of Weyl fermions). In the IR, the single  color-singlet $SU_f(2)$-fundamental composite ${\cal O}^{(2)\,\gamma}_{[ij]k}$ has a  a single zero mode  in the background of an $SU_f(2)$ instanton, which carries triple the charge of an elementary adjoint Weyl fermion under $\mathbb Z_4^{d\chi}$. In effect, the IR  $\mathbb Z_{4}^{d\chi}\left[SU_f(2)\right]^2$ anomaly  gives $3$, matching the UV anomaly.

To compute the gravitational anomaly  $\mathbb Z_{4}^{d\chi}\left[{\cal G}\right]^2$, we add up the contribution from all the flavor and color components (this counts zero modes in a gravitational instanton background, see \cite{Csaki:1997aw}). In the UV this gives $1\times 2\times 3=6$ ($1$ is the charge under $\mathbb Z_4^{d\chi}$, $2$ is the number of fundamental components and $3$ is $N_c^2-1$). In the IR this gives $3\times2\times 1=6$, where $3$ is the charge under $\mathbb Z_4^{d\chi}$. Again, we find an exact match between the UV and IR $\mathbb Z_{4}^{d\chi} \left[{\cal G}\right]^2$ anomaly.\footnote{We note that Ref.~\cite{Csaki:1997aw} also proposed matching the  less restrictive (as it is dependent on the massive spectrum) so-called ``type-II'' $\mathbb Z_4^3$ anomaly. It is easy to see that this condition is also obeyed here: $6\times 1^3 = 2 \times 3^3$(mod $m 4$), where $m \in \mathbb Z$, as per \cite{Csaki:1997aw}. }

Finally, we examine the recently discovered mixed $0$-form chiral/$1$-form center 't Hooft anomaly $\mathbb Z_{8}^{d\chi} \left[\mathbb Z_2^C\right]^2$ \cite{Gaiotto:2014kfa,Gaiotto:2017yup}. To see the anomaly, one has to introduce a $2$-form $\mathbb Z_N^C$ gauge field background for the center symmetry; it suffices to consider topological backgrounds (of zero $3$-form curvature). As discussed, e.g., in \cite{Komargodski:2017smk}, introducing topological  $\mathbb Z_2^C$ two-form background gauge fields is equivalent\footnote{This becomes especially clear---and is well-known---on the lattice, where the simplest $\mathbb Z_N^C$ two-form gauge field topological background is the one corresponding to the insertion of a thin center vortex (corresponding, e.g. to a $n_{12}$ 't Hooft flux in the continuum, see~Ch. 4 of \cite{Greensite:2011zz} and references therein).} to allowing nontrivial 't Hooft fluxes \cite{tHooft:1979rtg} when considering the theory on $\mathbb T^4$.

Now, under a discrete $ \mathbb Z_{8}^{d\chi}$ transformation, the fermion measure transforms by a phase factor $e^{ i 2 \pi Q}$. Here, $Q = {\int {\rm Tr} F \wedge F \over 8 \pi^2}$ is integer when the 't Hooft flux vanishes, hence, $\mathbb Z_{8}^{d\chi}$ is anomaly free in the theory without center-background gauge fields. However, $Q = {n_{12} n_{34}\over N_c} + n$, $n \in \mathbb Z$,  when, e.g., only the  't Hooft fluxes $n_{12}$ and $n_{34}$ (which are integers modulo $N_c$) are non-vanishing. Thus, for our two-color case, taking $n_{12}=n_{34}=1$,  the partition function acquires a phase $e^{i \pi}$ under $\mathbb Z_{8}^{d\chi}$  transformations. This phase is the manifestation of the mixed discrete-chiral/center-squared anomaly. It is independent of the $\mathbb T^4$ volume and can be argued to be renormalization group invariant: for example, it can  be seen to be due to the variation of a five dimensional  local  bulk  term involving only $\mathbb Z_{8}^{d\chi}$ and  and $\mathbb Z_2^C$ background fields \cite{Gaiotto:2014kfa,Gaiotto:2017yup,Komargodski:2017dmc,Komargodski:2017smk}.

 In the IR theory,  this anomaly is matched\footnote{\label{six}The observant reader may notice that the massless fermion (\ref{hadron}) also matches all purely $0$-form anomalies if the unbroken $\mathbb Z_4^{d \chi}$ chiral symmetry  is replaced by the full $\mathbb Z_8^{d \chi}$. However, an IR theory  of a single massless gauge-singlet fermion cannot match the mixed $0$-form/$1$-form discrete anomaly discussed here: it would not couple to 't Hooft flux and could not give rise to the mixed chiral/center anomaly, unless there were other massless (or topological) degrees of freedom present. Without such degrees of freedom,  the   
  $\mathbb Z_8^{d \chi} \rightarrow \mathbb Z_4^{d \chi}$ breaking appears as the only consistent possibility (see, however, \cite{Cordova:2018acb} for a scenario with unbroken $\mathbb Z_8^{d \chi}$ and an emergent massless gauge field). } in the ``Goldstone" (spontaneously broken) mode  due to the assumed nonvanishing ${\cal{O}}^{(1)}$ vacuum expectation value. A $\mathbb Z_{8}^{d\chi}$  symmetry transformation interchanges the two vacua and, in the background of 't Hooft fluxes, transforms the partition function by a $e^{i \pi}$ phase, as  in \cite{Gaiotto:2014kfa,Gaiotto:2017yup}.
 
As usual, anomaly-matching arguments do not  determine the spectrum of massive states associated  with the  discrete symmetry breaking. However, the recent insights of \cite{Gaiotto:2014kfa,Gaiotto:2017yup,Komargodski:2017dmc,Komargodski:2017smk} shed some light on the properties of the domain walls between the two vacua. For example, using a discrete version of anomaly inflow \cite{Gaiotto:2017yup}, one can argue that  the center symmetry---assumed or shown to be unbroken in the bulk, see below---is broken on the domain walls and the fundamental Wilson loop has a perimeter law there. This has been explicitly seen  in the semiclassical regime on $\mathbb R^3 \times \mathbb S^1$ \cite{Anber:2015kea} and also argued in \cite{Komargodski:2017smk}. It would be interesting to further study the dynamics of the various domain walls in this theory,  in the zero and finite temperature cases on both $\mathbb R^3 \times \mathbb S^1$  and $\mathbb R^4$  \cite{Shimizu:2017asf,Komargodski:2017smk,Anber:2011gn}. 
 
In summary, the 't Hooft matching conditions of $SU_c(2)$ Yang-Mills with two adjoint Weyl flavors allow for a phase with a massless composite fermion in the IR. The theory preserves its $SU_f(2)$ chiral symmetry and has two vacua that preserve a $\mathbb Z_4^{d\chi}$ subgroup of the $\mathbb Z_8^{d\chi}$ discrete chiral symmetry.

We also expect the theory to preserve its $\mathbb Z_2^{C}$ $1$-form symmetry, hence  the fundamental Wilson loop to obey an area law. The area-law expectation is consistent with the observed behavior as we compactify the theory on a circle $\mathbb S_L^1$ of circumference $L$ (where the fermions obey periodic boundary conditions along $\mathbb S_L^1$) and interpolate between $L\Lambda_{QCD}\ll1$ and $L\Lambda_{QCD}\gg1$. In the small circle limit one can integrate an infinite tower of Kaluza-Klein excitations of the gauge field and fermions along $\mathbb S_L^1$, which generates an effective potential for the holonomy $\Phi \equiv \oint_{\mathbb S_L^1} A$. This potential is minimized at a non-zero value of $\Phi$, which breaks $SU(2)$ down to $U(1)$.  Thus, the theory abelianizes, enters its weakly coupled regime (since $L\Lambda_{QCD}\ll1$), and  becomes amenable to semi-classical treatment. The abelian field can be taken along the $\tau^3$-direction (the Cartan component) in the color space, and thus, the adjoint fermions in this direction stays massless and uncharged\footnote{This can be easily seen from the covariant derivative of the adjoint fermions.} under $U(1)$.  Then, the effective three-dimensional $U(1)$ theory can be dualized to a free compact scalar $\sigma$. The story does not stop here since this theory admits magnetic bions \cite{Unsal:2007jx}, which are  non self-dual nonperturbative saddles.  The proliferation of magnetic bions leads to the generation of a potential, $V(\sigma)=\cos 2\sigma$; see \cite{Unsal:2007jx} for more details. Thus, the IR theory has two vacua at $\sigma=0$ and $\sigma=\pi$, and hence, the discrete chiral symmetry $\mathbb Z_{8}^{d\chi}$ breaks to $ \mathbb Z_{4}^{d\chi}$. This theory has a mass gap and confines the fundamental electric charges \cite{Unsal:2007jx}. In addition, as indicated above, there is an $SU_f(2)$-doublet massless excitation which is neutral under $U(1)$: the Cartan component of the adjoint fermions. Notice that this is the same number of massless fermions as in our proposed phase  on $\mathbb R^4$. Therefore, the theory in the semiclassical regime on $\mathbb R^3\times \mathbb S^1$ shares all the features of the proposed phase of the theory on $\mathbb R^4$ and we expect the continuity of the mass gap from the small to the large circle to hold as well.

{\flushleft\textbf{Discussion and relation to other studies:}} To recuperate our proposed spectrum and symmetries: we argue that in the $SU_c(2)$ theory with $n_f = 2$ massless Weyl adjoint fermions, the $SU_f(2)$ is unbroken, the theory has two vacua due to $\mathbb Z_8^{d \chi} \rightarrow \mathbb Z_4^{d \chi}$ symmetry breaking, and the IR spectrum consists of a single massless $SU_f(2)$-doublet Weyl fermion with the quantum numbers of Table \ref{charge table}.

For the theory at hand, we find it remarkable that the anomaly matching solution  on $\mathbb R^4$ realizes the IR spectrum and symmetries in a manner similar to the one found by semiclassically solving the theory on $\mathbb R^3 \times \mathbb S^1$. 

 A  question that may lurk in the reader's mind is whether such a continuity of the symmetry realization between small and large $\mathbb S^1$  is consistent with anomalies also in  higher rank and/or larger-$n_f$ QCD(adj) theories. While we cannot prove or disprove the existence of solutions of anomaly matching for a larger class of QCD(adj) theories, they are not trivial to construct. The case we discuss here is likely the simplest and  a devoted search for others is left for the future. 

As usual, any solution of 't Hooft's anomaly matching comes with a caveat---it only gives a possible consistent IR behavior of the theory, albeit one which may or may not be dynamically realized.
Clearly, to rule in or out our proposed realization of the symmetries, further analytical and numerical studies are needed. 

On the numerical side,  the $SU_c(2)$ gauge theory with a single Dirac flavor ($n_f = 2$ Weyl) in the adjoint has been the subject of the lattice studies \cite{Athenodorou:2014eua} and \cite{Bergner:2017gzw}, which found that the IR behavior of the theory appears inconsistent with the conventional continuous $SU_f(2) \rightarrow SO_f(2)$ chiral symmetry breaking, pointing ``tentatively" (the characterization is from \cite{Athenodorou:2014eua}) towards conformal or near conformal behavior. The authors considered only the two hitherto known possibilities consistent with anomaly matching: conformal behavior or continuous chiral symmetry breaking. The IR scenario we propose here offers a third way, but naturally poses challenges for lattice studies: the theory is, indeed, conformal in the deep IR, with the massless fermion the only relevant degree of freedom, but with a   likely complicated massive spectrum associated with the discrete symmetry breaking. The fermions are described by the Ginsparg-Wilson action with a bare mass (or hopping parameter), which will have the correct discrete chiral symmetry as well as the $0$-form and $1$-form mixed anomaly in the chiral limit. One can then study the histogram of the operator ${\cal O}^{(1)}$ as a function of the hopping parameter. In our scenario, the real part of this operator should fluctuate between two values in the continuum limit, indicating the breaking of $\mathbb Z^{d\chi}_{8}$ to $\mathbb Z_4^{d\chi}$. Some progress along this line of though in supersymmetric lattice models was achieved in \cite{Ali:2018dnd}.

On the analytical side, the beta function of YM theory with arbitrary representations   was recently computed to four loops \cite{Zoller:2016sgq}.
While the results are scheme dependent, it is amusing to note some interesting features  of the beta function. Adapting  \cite{Zoller:2016sgq} to the $n_f=2$ Weyl adjoint $SU_c(2)$ theory, one finds, first, the well-known result that  the  two-loop beta function shows no IR fixed point. Second, one finds that IR fixed points appear at three and four loops, with the value of the scheme dependent  fixed-point coupling becoming smaller as the number of loops is increased from three to four (we stress again that it is not obvious to us what these perturbative calculations teach us about the IR behavior of the theory). Another amusing exercise is to redo this  for other values of $n_f$. For $n_f = 1$ ($SU_c(2)$ super-Yang-Mills) one finds no IR fixed points up to the four loops of \cite{Zoller:2016sgq}, while for $n_f\ge 3$ a fixed point appears already at two loops (the fixed-point coupling moves to weaker values as the number of loops is increased from two to four). 

To rule  in or out conformality---or our proposal---one should  further improve the  lattice studies of  \cite{Athenodorou:2014eua} and  study other indications of conformality. These indications include: (i) the running coupling on the lattice in the $n_f=2$ theory,  as done for $n_f = 4$ in \cite{Rantaharju:2015yva} (we note that,  recently, \cite{Bergner:2017ytp} found no IR fixed point for $n_f = 2$ in the range of masses studied), (ii) the area vs. perimeter law for the Wilson loop, as advocated in \cite{Poppitz:2009uq}, and (iii) the expected nontrivial properties of domain walls, which should be present in any phase with broken chiral symmetry.

{\flushleft{Note added:}} While this letter was under review, the work \cite{Cordova:2018acb} appeared, showing that formulating our theory on a non-spin manifold gives rise to new 't Hooft anomalies, absent on spin manifolds (nonetheless, these anomalies impose constraints on the flat space theories). A consequence of these new anomalies is that our scenario for the IR physics  of two-flavor QCD(adj) should be supplemented with an  emergent $\mathbb Z_2$ gauge theory with  a broken (also emergent) $\mathbb Z_2$ 1-form global symmetry, which can be probed by appropriate 't Hooft loop operators, see \cite{Cordova:2018acb}.

{\flushleft\textbf{Acknowledgments:}} We thank Ken Intriligator and Julius Kuti for discussions relevant to this paper. We are also indebted to Thomas Dumitrescu for patiently explaining to us the results of \cite{Cordova:2018acb}. MA is supported by an NSF grant PHY-1720135 and the Murdock Charitable Trust. EP is supported by a Discovery Grant from NSERC.
\bibliographystyle{apsrev4-1}
\bibliography{QCD_adj_Anomalies_refs}

%merlin.mbs apsrev4-1.bst 2010-07-25 4.21a (PWD, AO, DPC) hacked
%Control: key (0)
%Control: author (72) initials jnrlst
%Control: editor formatted (1) identically to author
%Control: production of article title (-1) disabled
%Control: page (0) single
%Control: year (1) truncated
%Control: production of eprint (0) enabled
\begin{thebibliography}{22}%
\makeatletter
\providecommand \@ifxundefined [1]{%
 \@ifx{#1\undefined}
}%
\providecommand \@ifnum [1]{%
 \ifnum #1\expandafter \@firstoftwo
 \else \expandafter \@secondoftwo
 \fi
}%
\providecommand \@ifx [1]{%
 \ifx #1\expandafter \@firstoftwo
 \else \expandafter \@secondoftwo
 \fi
}%
\providecommand \natexlab [1]{#1}%
\providecommand \enquote  [1]{``#1''}%
\providecommand \bibnamefont  [1]{#1}%
\providecommand \bibfnamefont [1]{#1}%
\providecommand \citenamefont [1]{#1}%
\providecommand \href@noop [0]{\@secondoftwo}%
\providecommand \href [0]{\begingroup \@sanitize@url \@href}%
\providecommand \@href[1]{\@@startlink{#1}\@@href}%
\providecommand \@@href[1]{\endgroup#1\@@endlink}%
\providecommand \@sanitize@url [0]{\catcode `\\12\catcode `\$12\catcode
  `\&12\catcode `\#12\catcode `\^12\catcode `\_12\catcode `\%12\relax}%
\providecommand \@@startlink[1]{}%
\providecommand \@@endlink[0]{}%
\providecommand \url  [0]{\begingroup\@sanitize@url \@url }%
\providecommand \@url [1]{\endgroup\@href {#1}{\urlprefix }}%
\providecommand \urlprefix  [0]{URL }%
\providecommand \Eprint [0]{\href }%
\providecommand \doibase [0]{http://dx.doi.org/}%
\providecommand \selectlanguage [0]{\@gobble}%
\providecommand \bibinfo  [0]{\@secondoftwo}%
\providecommand \bibfield  [0]{\@secondoftwo}%
\providecommand \translation [1]{[#1]}%
\providecommand \BibitemOpen [0]{}%
\providecommand \bibitemStop [0]{}%
\providecommand \bibitemNoStop [0]{.\EOS\space}%
\providecommand \EOS [0]{\spacefactor3000\relax}%
\providecommand \BibitemShut  [1]{\csname bibitem#1\endcsname}%
\let\auto@bib@innerbib\@empty
%</preamble>
\bibitem [{\citenamefont {Unsal}(2009)}]{Unsal:2007jx}%
  \BibitemOpen
  \bibfield  {author} {\bibinfo {author} {\bibfnamefont {M.}~\bibnamefont
  {Unsal}},\ }\href {\doibase 10.1103/PhysRevD.80.065001} {\bibfield  {journal}
  {\bibinfo  {journal} {Phys. Rev.}\ }\textbf {\bibinfo {volume} {D80}},\
  \bibinfo {pages} {065001} (\bibinfo {year} {2009})},\ \Eprint
  {http://arxiv.org/abs/0709.3269} {arXiv:0709.3269 [hep-th]} \BibitemShut
  {NoStop}%
%%CITATION = ARXIV:0709.3269;%%
\bibitem [{\citenamefont {Dunne}\ and\ \citenamefont
  {Unsal}(2016)}]{Dunne:2016nmc}%
  \BibitemOpen
  \bibfield  {author} {\bibinfo {author} {\bibfnamefont {G.~V.}\ \bibnamefont
  {Dunne}}\ and\ \bibinfo {author} {\bibfnamefont {M.}~\bibnamefont {Unsal}},\
  }\href {\doibase 10.1146/annurev-nucl-102115-044755} {\bibfield  {journal}
  {\bibinfo  {journal} {Ann. Rev. Nucl. Part. Sci.}\ }\textbf {\bibinfo
  {volume} {66}},\ \bibinfo {pages} {245} (\bibinfo {year} {2016})},\ \Eprint
  {http://arxiv.org/abs/1601.03414} {arXiv:1601.03414 [hep-th]} \BibitemShut
  {NoStop}%
%%CITATION = ARXIV:1601.03414;%%
\bibitem [{\citenamefont {Frishman}\ \emph {et~al.}(1981)\citenamefont
  {Frishman}, \citenamefont {Schwimmer}, \citenamefont {Banks},\ and\
  \citenamefont {Yankielowicz}}]{Frishman:1980dq}%
  \BibitemOpen
  \bibfield  {author} {\bibinfo {author} {\bibfnamefont {Y.}~\bibnamefont
  {Frishman}}, \bibinfo {author} {\bibfnamefont {A.}~\bibnamefont {Schwimmer}},
  \bibinfo {author} {\bibfnamefont {T.}~\bibnamefont {Banks}}, \ and\ \bibinfo
  {author} {\bibfnamefont {S.}~\bibnamefont {Yankielowicz}},\ }\href {\doibase
  10.1016/0550-3213(81)90268-6} {\bibfield  {journal} {\bibinfo  {journal}
  {Nucl. Phys.}\ }\textbf {\bibinfo {volume} {B177}},\ \bibinfo {pages} {157}
  (\bibinfo {year} {1981})}\BibitemShut {NoStop}%
%%CITATION = NUPHA,B177,157;%%
\bibitem [{\citenamefont {Coleman}\ and\ \citenamefont
  {Grossman}(1982)}]{Coleman:1982yg}%
  \BibitemOpen
  \bibfield  {author} {\bibinfo {author} {\bibfnamefont {S.~R.}\ \bibnamefont
  {Coleman}}\ and\ \bibinfo {author} {\bibfnamefont {B.}~\bibnamefont
  {Grossman}},\ }\href {\doibase 10.1016/0550-3213(82)90028-1} {\bibfield
  {journal} {\bibinfo  {journal} {Nucl. Phys.}\ }\textbf {\bibinfo {volume}
  {B203}},\ \bibinfo {pages} {205} (\bibinfo {year} {1982})}\BibitemShut
  {NoStop}%
%%CITATION = NUPHA,B203,205;%%
\bibitem [{\citenamefont {Gaiotto}\ \emph {et~al.}(2015)\citenamefont
  {Gaiotto}, \citenamefont {Kapustin}, \citenamefont {Seiberg},\ and\
  \citenamefont {Willett}}]{Gaiotto:2014kfa}%
  \BibitemOpen
  \bibfield  {author} {\bibinfo {author} {\bibfnamefont {D.}~\bibnamefont
  {Gaiotto}}, \bibinfo {author} {\bibfnamefont {A.}~\bibnamefont {Kapustin}},
  \bibinfo {author} {\bibfnamefont {N.}~\bibnamefont {Seiberg}}, \ and\
  \bibinfo {author} {\bibfnamefont {B.}~\bibnamefont {Willett}},\ }\href
  {\doibase 10.1007/JHEP02(2015)172} {\bibfield  {journal} {\bibinfo  {journal}
  {JHEP}\ }\textbf {\bibinfo {volume} {02}},\ \bibinfo {pages} {172} (\bibinfo
  {year} {2015})},\ \Eprint {http://arxiv.org/abs/1412.5148} {arXiv:1412.5148
  [hep-th]} \BibitemShut {NoStop}%
%%CITATION = ARXIV:1412.5148;%%
\bibitem [{\citenamefont {Gaiotto}\ \emph {et~al.}(2017)\citenamefont
  {Gaiotto}, \citenamefont {Kapustin}, \citenamefont {Komargodski},\ and\
  \citenamefont {Seiberg}}]{Gaiotto:2017yup}%
  \BibitemOpen
  \bibfield  {author} {\bibinfo {author} {\bibfnamefont {D.}~\bibnamefont
  {Gaiotto}}, \bibinfo {author} {\bibfnamefont {A.}~\bibnamefont {Kapustin}},
  \bibinfo {author} {\bibfnamefont {Z.}~\bibnamefont {Komargodski}}, \ and\
  \bibinfo {author} {\bibfnamefont {N.}~\bibnamefont {Seiberg}},\ }\href
  {\doibase 10.1007/JHEP05(2017)091} {\bibfield  {journal} {\bibinfo  {journal}
  {JHEP}\ }\textbf {\bibinfo {volume} {05}},\ \bibinfo {pages} {091} (\bibinfo
  {year} {2017})},\ \Eprint {http://arxiv.org/abs/1703.00501} {arXiv:1703.00501
  [hep-th]} \BibitemShut {NoStop}%
%%CITATION = ARXIV:1703.00501;%%
\bibitem [{\citenamefont {Csaki}\ and\ \citenamefont
  {Murayama}(1998)}]{Csaki:1997aw}%
  \BibitemOpen
  \bibfield  {author} {\bibinfo {author} {\bibfnamefont {C.}~\bibnamefont
  {Csaki}}\ and\ \bibinfo {author} {\bibfnamefont {H.}~\bibnamefont
  {Murayama}},\ }\href {\doibase 10.1016/S0550-3213(97)00839-0} {\bibfield
  {journal} {\bibinfo  {journal} {Nucl. Phys.}\ }\textbf {\bibinfo {volume}
  {B515}},\ \bibinfo {pages} {114} (\bibinfo {year} {1998})},\ \Eprint
  {http://arxiv.org/abs/hep-th/9710105} {arXiv:hep-th/9710105 [hep-th]}
  \BibitemShut {NoStop}%
%%CITATION = HEP-TH/9710105;%%
\bibitem [{\citenamefont {Komargodski}\ \emph {et~al.}(2018)\citenamefont
  {Komargodski}, \citenamefont {Sulejmanpasic},\ and\ \citenamefont
  {Unsal}}]{Komargodski:2017smk}%
  \BibitemOpen
  \bibfield  {author} {\bibinfo {author} {\bibfnamefont {Z.}~\bibnamefont
  {Komargodski}}, \bibinfo {author} {\bibfnamefont {T.}~\bibnamefont
  {Sulejmanpasic}}, \ and\ \bibinfo {author} {\bibfnamefont {M.}~\bibnamefont
  {Unsal}},\ }\href {\doibase 10.1103/PhysRevB.97.054418} {\bibfield  {journal}
  {\bibinfo  {journal} {Phys. Rev.}\ }\textbf {\bibinfo {volume} {B97}},\
  \bibinfo {pages} {054418} (\bibinfo {year} {2018})},\ \Eprint
  {http://arxiv.org/abs/1706.05731} {arXiv:1706.05731 [cond-mat.str-el]}
  \BibitemShut {NoStop}%
%%CITATION = ARXIV:1706.05731;%%
\bibitem [{\citenamefont {Greensite}(2011)}]{Greensite:2011zz}%
  \BibitemOpen
  \bibfield  {author} {\bibinfo {author} {\bibfnamefont {J.}~\bibnamefont
  {Greensite}},\ }\href {\doibase 10.1007/978-3-642-14382-3} {\bibfield
  {journal} {\bibinfo  {journal} {Lect. Notes Phys.}\ }\textbf {\bibinfo
  {volume} {821}},\ \bibinfo {pages} {1} (\bibinfo {year} {2011})}\BibitemShut
  {NoStop}%
%%CITATION = LNPHA,821,1;%%
\bibitem [{\citenamefont {'t~Hooft}(1979)}]{tHooft:1979rtg}%
  \BibitemOpen
  \bibfield  {author} {\bibinfo {author} {\bibfnamefont {G.}~\bibnamefont
  {'t~Hooft}},\ }\href {\doibase 10.1016/0550-3213(79)90595-9} {\bibfield
  {journal} {\bibinfo  {journal} {Nucl. Phys.}\ }\textbf {\bibinfo {volume}
  {B153}},\ \bibinfo {pages} {141} (\bibinfo {year} {1979})}\BibitemShut
  {NoStop}%
%%CITATION = NUPHA,B153,141;%%
\bibitem [{\citenamefont {Komargodski}\ \emph {et~al.}(2017)\citenamefont
  {Komargodski}, \citenamefont {Sharon}, \citenamefont {Thorngren},\ and\
  \citenamefont {Zhou}}]{Komargodski:2017dmc}%
  \BibitemOpen
  \bibfield  {author} {\bibinfo {author} {\bibfnamefont {Z.}~\bibnamefont
  {Komargodski}}, \bibinfo {author} {\bibfnamefont {A.}~\bibnamefont {Sharon}},
  \bibinfo {author} {\bibfnamefont {R.}~\bibnamefont {Thorngren}}, \ and\
  \bibinfo {author} {\bibfnamefont {X.}~\bibnamefont {Zhou}},\ }\href@noop {}
  {\  (\bibinfo {year} {2017})},\ \Eprint {http://arxiv.org/abs/1705.04786}
  {arXiv:1705.04786 [hep-th]} \BibitemShut {NoStop}%
%%CITATION = ARXIV:1705.04786;%%
\bibitem [{\citenamefont {Cordova}\ and\ \citenamefont
  {Dumitrescu}(2018)}]{Cordova:2018acb}%
  \BibitemOpen
  \bibfield  {author} {\bibinfo {author} {\bibfnamefont {C.}~\bibnamefont
  {Cordova}}\ and\ \bibinfo {author} {\bibfnamefont {T.~T.}\ \bibnamefont
  {Dumitrescu}},\ }\href@noop {} {\  (\bibinfo {year} {2018})},\ \Eprint
  {http://arxiv.org/abs/1806.09592} {arXiv:1806.09592 [hep-th]} \BibitemShut
  {NoStop}%
%%CITATION = ARXIV:1806.09592;%%
\bibitem [{\citenamefont {Anber}\ \emph {et~al.}(2015)\citenamefont {Anber},
  \citenamefont {Poppitz},\ and\ \citenamefont
  {Sulejmanpasic}}]{Anber:2015kea}%
  \BibitemOpen
  \bibfield  {author} {\bibinfo {author} {\bibfnamefont {M.~M.}\ \bibnamefont
  {Anber}}, \bibinfo {author} {\bibfnamefont {E.}~\bibnamefont {Poppitz}}, \
  and\ \bibinfo {author} {\bibfnamefont {T.}~\bibnamefont {Sulejmanpasic}},\
  }\href {\doibase 10.1103/PhysRevD.92.021701} {\bibfield  {journal} {\bibinfo
  {journal} {Phys. Rev.}\ }\textbf {\bibinfo {volume} {D92}},\ \bibinfo {pages}
  {021701} (\bibinfo {year} {2015})},\ \Eprint
  {http://arxiv.org/abs/1501.06773} {arXiv:1501.06773 [hep-th]} \BibitemShut
  {NoStop}%
%%CITATION = ARXIV:1501.06773;%%
\bibitem [{\citenamefont {Shimizu}\ and\ \citenamefont
  {Yonekura}(2018)}]{Shimizu:2017asf}%
  \BibitemOpen
  \bibfield  {author} {\bibinfo {author} {\bibfnamefont {H.}~\bibnamefont
  {Shimizu}}\ and\ \bibinfo {author} {\bibfnamefont {K.}~\bibnamefont
  {Yonekura}},\ }\href {\doibase 10.1103/PhysRevD.97.105011} {\bibfield
  {journal} {\bibinfo  {journal} {Phys. Rev.}\ }\textbf {\bibinfo {volume}
  {D97}},\ \bibinfo {pages} {105011} (\bibinfo {year} {2018})},\ \Eprint
  {http://arxiv.org/abs/1706.06104} {arXiv:1706.06104 [hep-th]} \BibitemShut
  {NoStop}%
%%CITATION = ARXIV:1706.06104;%%
\bibitem [{\citenamefont {Anber}\ \emph {et~al.}(2012)\citenamefont {Anber},
  \citenamefont {Poppitz},\ and\ \citenamefont {Unsal}}]{Anber:2011gn}%
  \BibitemOpen
  \bibfield  {author} {\bibinfo {author} {\bibfnamefont {M.~M.}\ \bibnamefont
  {Anber}}, \bibinfo {author} {\bibfnamefont {E.}~\bibnamefont {Poppitz}}, \
  and\ \bibinfo {author} {\bibfnamefont {M.}~\bibnamefont {Unsal}},\ }\href
  {\doibase 10.1007/JHEP04(2012)040} {\bibfield  {journal} {\bibinfo  {journal}
  {JHEP}\ }\textbf {\bibinfo {volume} {04}},\ \bibinfo {pages} {040} (\bibinfo
  {year} {2012})},\ \Eprint {http://arxiv.org/abs/1112.6389} {arXiv:1112.6389
  [hep-th]} \BibitemShut {NoStop}%
%%CITATION = ARXIV:1112.6389;%%
\bibitem [{\citenamefont {Athenodorou}\ \emph {et~al.}(2015)\citenamefont
  {Athenodorou}, \citenamefont {Bennett}, \citenamefont {Bergner},\ and\
  \citenamefont {Lucini}}]{Athenodorou:2014eua}%
  \BibitemOpen
  \bibfield  {author} {\bibinfo {author} {\bibfnamefont {A.}~\bibnamefont
  {Athenodorou}}, \bibinfo {author} {\bibfnamefont {E.}~\bibnamefont
  {Bennett}}, \bibinfo {author} {\bibfnamefont {G.}~\bibnamefont {Bergner}}, \
  and\ \bibinfo {author} {\bibfnamefont {B.}~\bibnamefont {Lucini}},\ }\href
  {\doibase 10.1103/PhysRevD.91.114508} {\bibfield  {journal} {\bibinfo
  {journal} {Phys. Rev.}\ }\textbf {\bibinfo {volume} {D91}},\ \bibinfo {pages}
  {114508} (\bibinfo {year} {2015})},\ \Eprint {http://arxiv.org/abs/1412.5994}
  {arXiv:1412.5994 [hep-lat]} \BibitemShut {NoStop}%
%%CITATION = ARXIV:1412.5994;%%
\bibitem [{\citenamefont {Bergner}\ \emph {et~al.}(2018)\citenamefont
  {Bergner}, \citenamefont {Giudice}, \citenamefont {MŸnster}, \citenamefont
  {Scior}, \citenamefont {Montvay},\ and\ \citenamefont
  {Piemonte}}]{Bergner:2017gzw}%
  \BibitemOpen
  \bibfield  {author} {\bibinfo {author} {\bibfnamefont {G.}~\bibnamefont
  {Bergner}}, \bibinfo {author} {\bibfnamefont {P.}~\bibnamefont {Giudice}},
  \bibinfo {author} {\bibfnamefont {G.}~\bibnamefont {MŸnster}}, \bibinfo
  {author} {\bibfnamefont {P.}~\bibnamefont {Scior}}, \bibinfo {author}
  {\bibfnamefont {I.}~\bibnamefont {Montvay}}, \ and\ \bibinfo {author}
  {\bibfnamefont {S.}~\bibnamefont {Piemonte}},\ }\href {\doibase
  10.1007/JHEP01(2018)119} {\bibfield  {journal} {\bibinfo  {journal} {JHEP}\
  }\textbf {\bibinfo {volume} {01}},\ \bibinfo {pages} {119} (\bibinfo {year}
  {2018})},\ \Eprint {http://arxiv.org/abs/1712.04692} {arXiv:1712.04692
  [hep-lat]} \BibitemShut {NoStop}%
%%CITATION = ARXIV:1712.04692;%%
\bibitem [{\citenamefont {Ali}\ \emph {et~al.}(2018)\citenamefont {Ali},
  \citenamefont {Bergner}, \citenamefont {Gerber}, \citenamefont {Giudice},
  \citenamefont {Montvay}, \citenamefont {Münster}, \citenamefont {Piemonte},\
  and\ \citenamefont {Scior}}]{Ali:2018dnd}%
  \BibitemOpen
  \bibfield  {author} {\bibinfo {author} {\bibfnamefont {S.}~\bibnamefont
  {Ali}}, \bibinfo {author} {\bibfnamefont {G.}~\bibnamefont {Bergner}},
  \bibinfo {author} {\bibfnamefont {H.}~\bibnamefont {Gerber}}, \bibinfo
  {author} {\bibfnamefont {P.}~\bibnamefont {Giudice}}, \bibinfo {author}
  {\bibfnamefont {I.}~\bibnamefont {Montvay}}, \bibinfo {author} {\bibfnamefont
  {G.}~\bibnamefont {Münster}}, \bibinfo {author} {\bibfnamefont
  {S.}~\bibnamefont {Piemonte}}, \ and\ \bibinfo {author} {\bibfnamefont
  {P.}~\bibnamefont {Scior}},\ }\href {\doibase 10.1007/JHEP03(2018)113}
  {\bibfield  {journal} {\bibinfo  {journal} {JHEP}\ }\textbf {\bibinfo
  {volume} {03}},\ \bibinfo {pages} {113} (\bibinfo {year} {2018})},\ \Eprint
  {http://arxiv.org/abs/1801.08062} {arXiv:1801.08062 [hep-lat]} \BibitemShut
  {NoStop}%
%%CITATION = ARXIV:1801.08062;%%
\bibitem [{\citenamefont {Zoller}(2016)}]{Zoller:2016sgq}%
  \BibitemOpen
  \bibfield  {author} {\bibinfo {author} {\bibfnamefont {M.~F.}\ \bibnamefont
  {Zoller}},\ }\href {\doibase 10.1007/JHEP10(2016)118} {\bibfield  {journal}
  {\bibinfo  {journal} {JHEP}\ }\textbf {\bibinfo {volume} {10}},\ \bibinfo
  {pages} {118} (\bibinfo {year} {2016})},\ \Eprint
  {http://arxiv.org/abs/1608.08982} {arXiv:1608.08982 [hep-ph]} \BibitemShut
  {NoStop}%
%%CITATION = ARXIV:1608.08982;%%
\bibitem [{\citenamefont {Rantaharju}\ \emph {et~al.}(2016)\citenamefont
  {Rantaharju}, \citenamefont {Rantalaiho}, \citenamefont {Rummukainen},\ and\
  \citenamefont {Tuominen}}]{Rantaharju:2015yva}%
  \BibitemOpen
  \bibfield  {author} {\bibinfo {author} {\bibfnamefont {J.}~\bibnamefont
  {Rantaharju}}, \bibinfo {author} {\bibfnamefont {T.}~\bibnamefont
  {Rantalaiho}}, \bibinfo {author} {\bibfnamefont {K.}~\bibnamefont
  {Rummukainen}}, \ and\ \bibinfo {author} {\bibfnamefont {K.}~\bibnamefont
  {Tuominen}},\ }\href {\doibase 10.1103/PhysRevD.93.094509} {\bibfield
  {journal} {\bibinfo  {journal} {Phys. Rev.}\ }\textbf {\bibinfo {volume}
  {D93}},\ \bibinfo {pages} {094509} (\bibinfo {year} {2016})},\ \Eprint
  {http://arxiv.org/abs/1510.03335} {arXiv:1510.03335 [hep-lat]} \BibitemShut
  {NoStop}%
%%CITATION = ARXIV:1510.03335;%%
\bibitem [{\citenamefont {Bergner}\ and\ \citenamefont
  {Piemonte}(2018)}]{Bergner:2017ytp}%
  \BibitemOpen
  \bibfield  {author} {\bibinfo {author} {\bibfnamefont {G.}~\bibnamefont
  {Bergner}}\ and\ \bibinfo {author} {\bibfnamefont {S.}~\bibnamefont
  {Piemonte}},\ }\href {\doibase 10.1103/PhysRevD.97.074510} {\bibfield
  {journal} {\bibinfo  {journal} {Phys. Rev.}\ }\textbf {\bibinfo {volume}
  {D97}},\ \bibinfo {pages} {074510} (\bibinfo {year} {2018})},\ \Eprint
  {http://arxiv.org/abs/1709.07367} {arXiv:1709.07367 [hep-lat]} \BibitemShut
  {NoStop}%
%%CITATION = ARXIV:1709.07367;%%
\bibitem [{\citenamefont {Poppitz}\ and\ \citenamefont
  {Unsal}(2009)}]{Poppitz:2009uq}%
  \BibitemOpen
  \bibfield  {author} {\bibinfo {author} {\bibfnamefont {E.}~\bibnamefont
  {Poppitz}}\ and\ \bibinfo {author} {\bibfnamefont {M.}~\bibnamefont
  {Unsal}},\ }\href {\doibase 10.1088/1126-6708/2009/09/050} {\bibfield
  {journal} {\bibinfo  {journal} {JHEP}\ }\textbf {\bibinfo {volume} {09}},\
  \bibinfo {pages} {050} (\bibinfo {year} {2009})},\ \Eprint
  {http://arxiv.org/abs/0906.5156} {arXiv:0906.5156 [hep-th]} \BibitemShut
  {NoStop}%
%%CITATION = ARXIV:0906.5156;%%
\end{thebibliography}%

\end{document}